# Enhancing Realism in Holographic Augmented Reality Displays through Occlusion Handling

*Woongseob Han, Chanseul Lee, and Jae-Hyeung Park**

W. Han, C. Lee, and J.-H. Park

Department of Electrical and Computer Engineering, Seoul National University, Seoul, South Korea

E-mail: J.-H. Park (jaehyeung@snu.ac.kr)

Co-author's E-mail: W. Han (dndtjq89@snu.ac.kr), C. Lee (lcs8913@snu.ac.kr).

Funding: National Research Foundation of Korea (NRF) [Grant No.RS-2022-NR070432, 34%]; [Grant No. RS-2024-00416272, 33%]; and [Grant No. RS-2024-00414230, 33%].

## Abstract

In this paper, an occlusion-capable holographic augmented-reality (AR) display is proposed, and its ability to enhance AR imagery through occlusion is demonstrated. Holographic displays can generate ideal three-dimensional (3D) virtual images and have recently shown rapid advancements, particularly in noise reduction through learning-based approaches. However, these displays still face challenges in improving image quality for AR scenarios because holographic virtual images are simply superimposed onto the real world, leading to a loss of contrast and visibility. To address this, an occlusion optics, which can mask designated areas of the real world, is incorporated into holographic AR displays. The proposed system employs a folded 4f system with a digital micromirror device and sequentially operates as both a real-world mask and an active Fourier filter. This approach transforms traditionally translucent holographic images into perceptually opaque ones while simultaneously eliminating unwanted noise terms from pixelated holographic displays. Furthermore, active Fourier filtering expands the virtual image field of view through time-multiplexed operation and supports a novel binary hologram optimization algorithm that performs especially well for sparse virtual content. The implementation successfully achieves opaque holographic 3D image presentation, significantly improving contrast and image quality while producing highly realistic 3D AR scenes with optically cast shadows.

## 1. Introduction

Augmented reality (AR) is a combination of real environments and virtual information designed to assist users in various aspects of daily life, including navigation, entertainment, and education.[1, 2] To visualize AR environments, two types of AR displays have been developed: video see-through (VST), also known as digital pass-through displays, and optical see-through (OST) displays.[3] VST displays capture the real-world through cameras and computationally merge it with virtual information to create the AR scene, allowing the user to view the real world through the display. In contrast, OST displays present AR environments by using a transparent optical combiner that optically superimposes virtual information onto the real scene.[4] Since the real-world is directly observed through the transparent combiner, OST displays are often regarded as the ultimate goal. However, imperfect visual representation of virtual information in conventional OST displays hinders their broader adoption. These visual defects typically manifest as unnatural three-dimensional (3D) imagery and contrast-impaired translucent images. The former issue stems from the lack of the accommodation cue, making holography, which reconstructs the wavefront of light with full focal cues, an ideal solution.[5, 6] The latter issue results from the ambient light from the real-world. To address this, additional occlusion optics capable of blocking the light from the real objects must be integrated into AR devices.[7]

Occlusion refers to the phenomenon where objects in the foreground obscure those in the background.[8] Occlusion between real objects is determined by the inherent material properties of the foreground object. Depending on the transparency and roughness of the foreground material, light from the background object may be reflected, scattered, or transmitted, resulting in either opaque or transparent imagery. However, in conventional OST displays, virtual images are optically formed in the air, failing to interact with light from the background.[9] Background light from the real world consistently passes through the foreground virtual images, degrading their contrast. This limitation also leads to depth-ambiguous AR presentations, as occlusion serves as one of the most powerful depth cues in human vision. The same limitation applies to holographic displays, which are capable of delivering high-resolution images with a continuous depth spectrum.[10] Although learning-based approaches to computer-generated hologram (CGH) synthesis have accelerated advancements in holographic displays—particularly in high-quality 3D image presentation[11-13]—AR scenarios still suffer from the impact of background light. Therefore, for AR devices, an additional occlusion system capable of interacting with background light from the real world should be adopted.

Implementing occlusion in AR displays varies in complexity depending on the occlusion method employed. One approach, known as the soft-edge occlusion, simply places the spatial light modulator (SLM) directly in front of the eye to block the light from the real world.[14-16] Due to its simplicity and compact system design, this method has already been implemented in the commercial product.[17] However, soft-edge occlusion always presents blurry mask as the human eye focuses on objects located at a distance, whereas the mask is positioned very close to the eye, making it difficult to bring the mask into clear focus. An alternative method for achieving ideal occlusion is hard-edge occlusion. This method utilizes a 4f system, where the SLM is positioned within the setup to generate an active mask at a specific depth.[18-22] By aligning the transverse and depth position of the virtual image with that of the real-scene mask, the virtual image is perceived as a physical object that optically interacts with the real world. Leveraging this interaction with real-world light, the system can control transparency and contrast, and even create artificial shadows.[23, 24]

Since being pioneered by Dennis Gabor,[25] holography has become both an ultimate goal and a long-standing challenge for researchers in related fields. Especially, dynamic holograms—commonly referred to as holographic displays—continue to face challenges such as the limited space bandwidth product (SBP) of SLMs and speckle noise in holographic reconstructions. Furthermore, generating CGHs from intensity-only targets, also known as the phase retrieval problem, is an ill-posed challenge, making the synthesis of optimized holograms difficult. Notably, recent progress in gradient descent algorithms and deep learning technologies has substantially contributed to addressing phase retrieval and speckle noise in layer-based CGH synthesis, thereby enhancing the quality of reconstructed holographic images.[12, 26, 27]

For clear holographic presentations, however, it is important to reduce other types of noise such as high-order, DC, and in case of amplitude holograms, conjugate noise. Conventional holographic displays utilize a 4f system with a static Fourier filter to eliminate these noises. Although recent studies on phase holograms demonstrated the feasibility of operation without a Fourier filter,[28-30] suppressing DC and high-order noise solely through CGH synthesis imposes significant loads on the SLM, making it challenging to generate high quality 3D presentations. In amplitude holograms, single-sideband (SSB) filtering,[31-33] which effectively suppresses conjugate noise, has been widely adopted. However, the SSB technique stops half of spatial frequency to remove the conjugate, thereby limiting SBP and field of view (FoV).[34]

In this paper, we propose a holographic OST-AR display that enhances the visual realism in AR presentation by handling occlusion. The proposed method integrates hard-edge occlusion optics into a holographic display system by replacing the static Fourier filter with an active amplitude SLM. Positioned in the Fourier plane of a 4f optical system, the amplitude SLM serves dual roles: it functions as an occlusion mask for the real-world see-through view and as an active Fourier filter for the holographic display, bringing benefits to both aspects of the system. The occlusion mask produces a hard-edged region aligned with the holographic virtual image by selectively blocking real-world light from designated areas, thereby enhancing the contrast and visibility of the holographic images. This mask not only blocks real-world light behind the virtual images but also allows for casting shadows of virtual objects onto the real scene, further boosting AR realism. In addition, the active Fourier filter maximizes the SBP usage of the SLM, effectively doubling the FoV compared to static filter-based systems. It also enables non-conventional CGH optimization, leading to superior holographic reconstruction quality relative to conventional static-filter-based CGHs. To support this, we introduce a novel binary CGH (bCGH) algorithm, specifically optimized for AR scenarios with the active filter. For implementation, we employed two high-refresh-rate binary SLMs—one for holographic display and one for the Fourier filter and occlusion mask—synchronized with a laser light source. The high refresh rates of the SLMs also allow for speckle suppression via time-multiplexed holographic reconstruction, further enhancing image quality. Optical experiments validate that the proposed holographic OST-AR display with occlusion support can generate high-contrast, high-quality 3D AR scenes, including artificially cast shadows onto the real world.

## 2. Results

### 2.1. System Architecture and operation

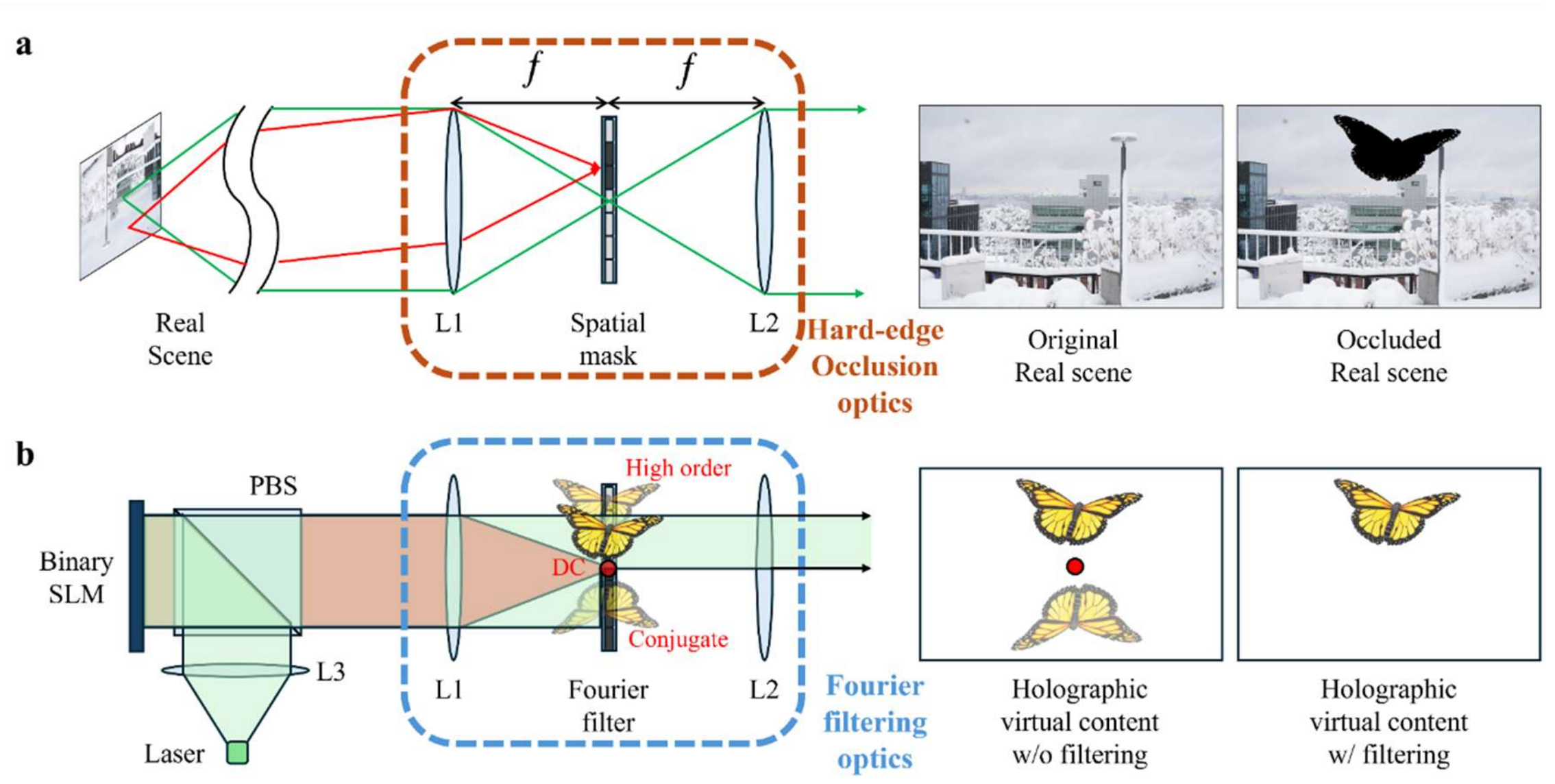

**Figure 1.** Simplified optics of the proposed system. The 4f system for a) hard-edge occlusion of the real scene and b) Fourier filtering in holographic displays is systematically identical, enabling both real scene occlusion and clear holographic reconstruction using a single 4f system.

Our optical system design is inspired by the structural similarity between hard-edge occlusion optics and Fourier filtering optics. Both systems utilize 4f optics with a spatial mask, allowing a single 4f optics to simultaneously serve as both an occlusion and a filtering system. **Figure 1** illustrates the simplified optical configurations of each system integrated into the proposed method. The hard-edge occlusion optics is depicted in Figure 1a. In this setup, a 4f system creates an intermediate plane of the real world between two convex lenses. By placing a pixelated spatial mask at this intermediate plane, the depth position of the mask can be aligned with a specific depth in the real scene. For instance, when the spatial mask is positioned at the Fourier plane of the 4f system as shown in Figure 1a, the optical output represents the real scene occluded by a mask at optical infinity.

Figure 1b shows the simplified Fourier filtering optics incorporated into the proposed system. In this configuration, a binary-amplitude SLM is used to generate speckle-suppressed holographic images by displaying multiple bCGHs within the eye integration time (<1/30s). To suppress DC, high-order, and conjugate noise, the system employs a pixelated active Fourier filter. This filter selectively passes the holographic image areas while blocking other frequency components, enabling noise-free holographic reconstructions.

Notably, the active pixelated Fourier filtering scheme offers two additional advantages over conventional static filtering approaches. First, it enables maximal utilization of the SLM's SBP. Traditional SSB amplitude holograms lose half of the SBP to filter out conjugate noise. In contrast, the proposed system fully utilizes the SLM's SBP by time-sequentially controlling the active Fourier filter. Specifically, in the first frame, one half of the Fourier plane is opened as a region of interest (ROI) containing holographic content, while the other half is closed to block conjugate noise. In the next frame, the roles are reversed: the previously closed side becomes the ROI and the former ROI area is closed. This approach effectively doubles the FoV of the proposed system.

Second, the active pixelated Fourier filter improves holographic image quality. Unlike non-pixelated filters that simply open one half of the Fourier plane, the pixelated filter selectively opens only the areas containing holographic content. This selective filtering enlarges the solution space for the bCGH optimization algorithm based on stochastic gradient decent (SGD), allowing the optimizer to disregard the masked regions and focus exclusively on the image regions. As a result, the active pixelated Fourier filter-based bCGH algorithm achieves greater loss reduction compared to conventional methods, particularly when generating sparse holographic content. For more details on the proposed binary hologram synthesis algorithm, refer to Section 2.2.

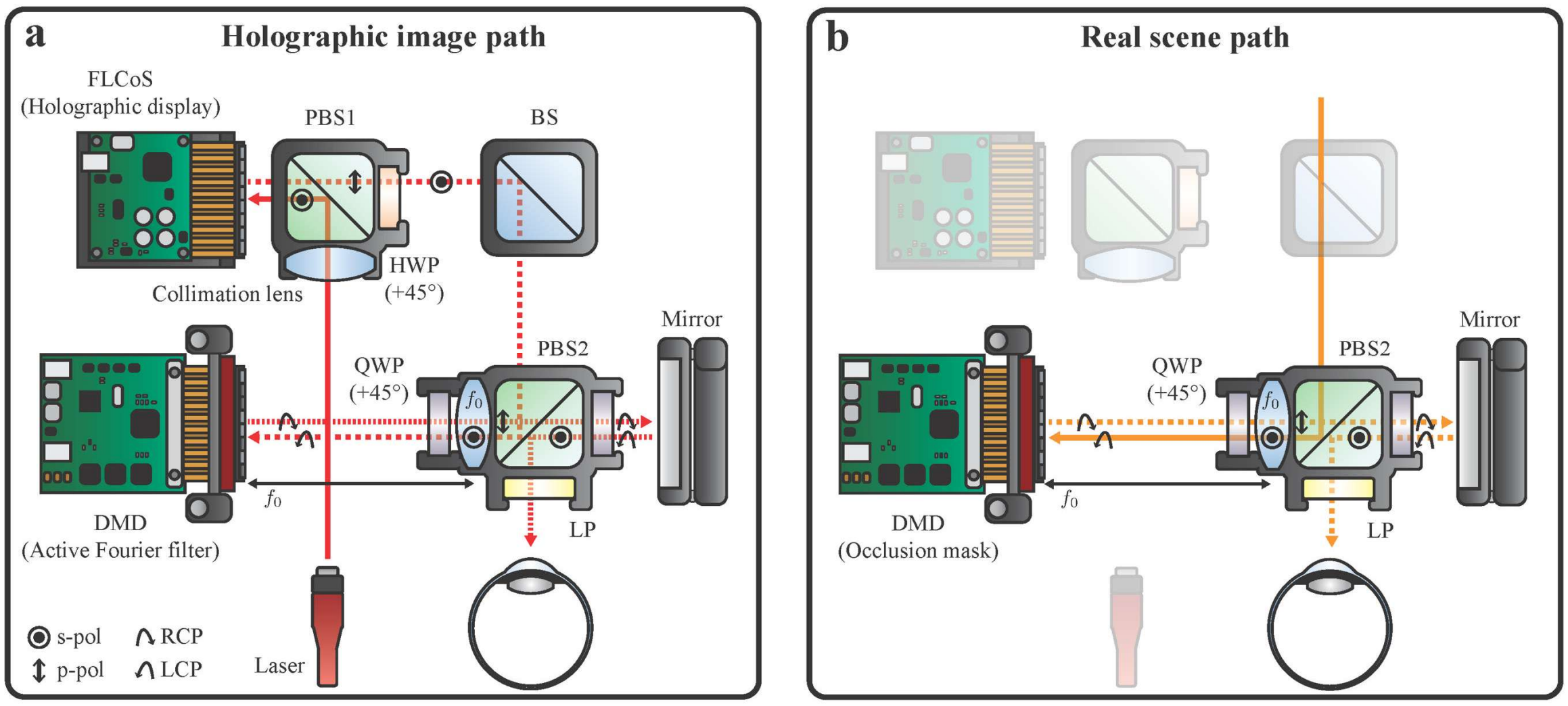

**Figure 2.** Schematic diagrams of the proposed method depicting optical paths of a) holographic image and b) real scene. A DMD is sequentially operated as a real scene mask and a Fourier filter to create immersive AR imagery, while a FLCoS works as a binary holographic display.

Since the structure of both systems is identical, as shown in Figure 1, we utilize a single folded 4f optics, operating it time-sequentially as both occlusion and Fourier filtering optics. **Figure 2** illustrates the complete structure of the proposed system, including the optical paths for both holographic images and the real world. On the top side, the components such as a ferroelectric liquid crystal on silicon (FLCoS), a polarization beam splitter (PBS) with a collimation lens, a half-wave plate (HWP), and a BS, generate holographic images and direct them to the Fourier filtering optics on the bottom side. Specifically, diverging light from a fiber laser is collimated, and only the s-polarized component is reflected toward the FLCoS by the PBS. The FLCoS, which displays bCGHs, modulates the wavefront of the light and transforms its polarization state to p-polarization, thereby generating holographic content along with other noise as it passes through the PBS1. The output light is then converted back to s-polarization by the HWP, reflected by the BS, and guided toward the bottom side of the system.

At the bottom side, the folded 4f optics with a quarter-wave plate (QWP) for occlusion and Fourier filtering is positioned to the left of PBS2, while an additional folding optics, consisting of another QWP and a mirror, is located to the right. Light from top-side, containing the holographic image and other noise components, travels toward the PBS2, is reflected, and enters the folded 4f system. Here, the light is refracted by the collimation lens with a focal length of $f_o$, propagates through free space over a distance of $f_o$, and reaches the Fourier plane. A digital micromirror device (DMD) at the Fourier plane functions as an active Fourier filter, suppressing the conjugate, high order, and DC noise components. As a result, only the desired holographic images are reflected back and exits the 4f optics. The QWP within the 4f optics ensures the output light is converted to p-polarized light, allowing it to pass through the PBS2 instead of being reflected. Finally, the light is reflected back by the mirror on the right and by PBS2, being directed toward the user.

The light from the real world interacts only with the bottom-side components, following an optical path identical to that of the holographic image. However, unlike in the holographic imaging case, the DMD functions as an occlusion mask, blocking the real world at optical infinity. As the occlusion mask exhibits a pattern different from the Fourier filter, time-multiplexed operation with mask sub-frame (MSF) and image sub-frame (ISF) is used (See Section 2.3 for detailed operation strategy). To achieve this, the FLCoS, DMD, and fiber laser are synchronized using an Arduino. For more details on the synchronization process, refer to Section S2 (Supporting Information).

### 2.2. Active filter-based binary hologram synthesis

In this section, we present our active filter-based bCGH synthesis method. Built on Fourier holography, our approach reconstructs holographic images near the Fourier plane to fully leverage the capabilities of the active Fourier filter. Following the recent study on bCGH algorithm,[35] we adopt an SGD-based time-multiplexed bCGH synthesis combined with SSB encoding to achieve speckle-reduced holographic reconstruction. The primary distinction of our method compared to the original learning-based bCGH algorithm is the integration of an active Fourier filter, which enables superior loss reduction in AR scenarios featuring a full-FoV real-world environment with sparse virtual content. The pipeline of our algorithm is shown in **Figure 3.**

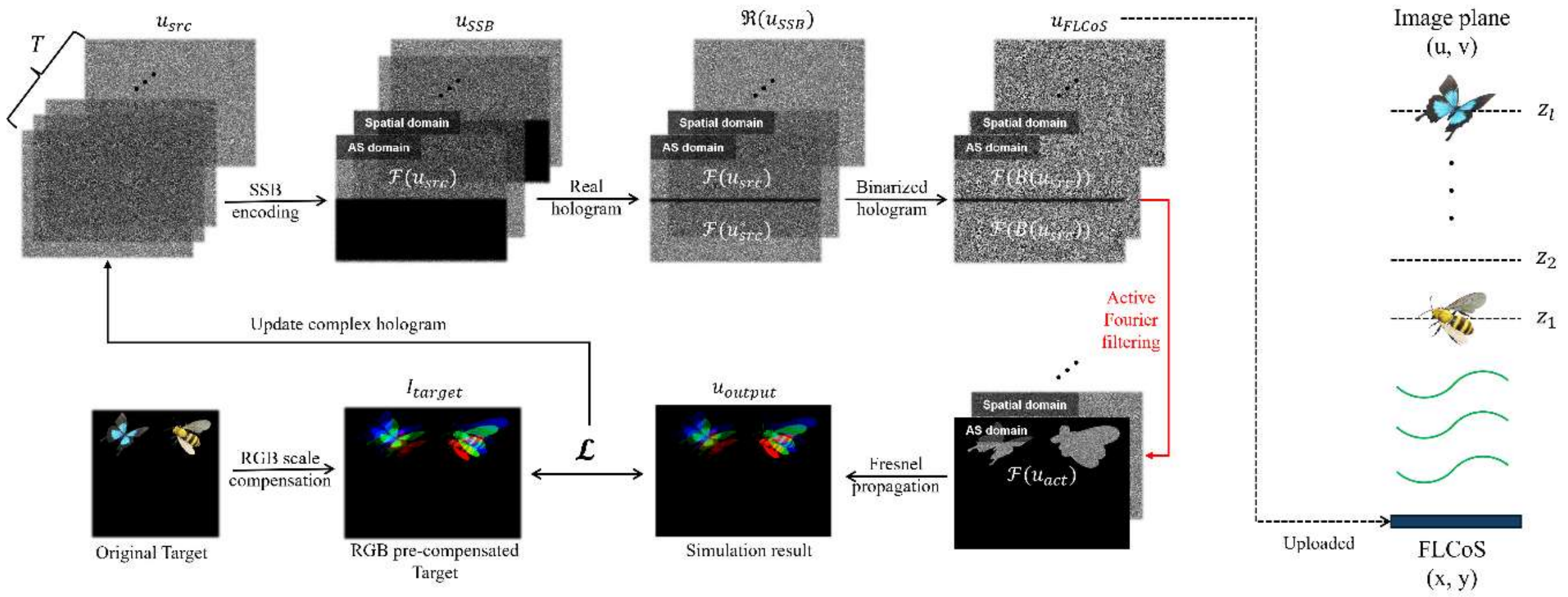

**Figure 3.** Algorithm pipeline. The initial complex hologram undergoes SSB encoding, is converted to real and binary form, content-dependently Fourier filtered, and propagated using a single FFT Fresnel propagation. The pre-compensated target image for full-color matching is used to calculate the loss and update the complex hologram.

As the FLCoS is used as a holographic display in the proposed system, we establish an SGD-based iterative algorithm to generate bCGH. The initial complex hologram $u_{src}$ is defined as

$$u_{src}(x,y) = a(x,y) \cdot e^{j\phi(x,y)}, \tag{1}$$

where $a(x,y)$and $\phi(x,y)$ are amplitude and phase of the complex hologram, respectively. Next, we apply SSB technique as follows.

$$u_{SSB}(x,y) = \mathcal{F}^{-1}\{\mathcal{F}\{u_{src}(x,y)\} \cdot H_{SSB}(\xi,\eta)\}, \tag{2}$$

$$H_{SSB}(\xi,\eta)=\begin{cases}1, & if\ \eta>0,\\ 0, & if\ \eta\leq 0,\end{cases} \tag{3}$$

where $\mathcal{F}\{\cdot\}$ is the Fourier transform operator and $H_{SSB}(\xi,\eta)$ is the kernel of the SSB filter in the Fourier domain$(\xi,\eta)$. To generate the binary-amplitude CGH uploaded on FLCoS, we take only the real part of the filtered complex hologram $u_{SSB}$ and binarize it as,

$$u_{FLCoS}(x,y)=B\{\mathrm{R}\{u_{SSB}(x,y)\}\}, \tag{4}$$

where R is real part operator and $B=sign(x,y)$ is a binarization operator. The resultant bCGH $u_{FLCoS}$ is displayed on FLCoS. Note that this simple thresholding binarization is non-differentiable, which prohibits gradient-based optimization. Therefore, we replace the binarization operator with a differentiable sigmoid function during the back-propagation process to enable gradient computation. Next, we simulate the active Fourier filtering as,

$$u_{act}(x,y)=\mathcal{F}^{-1}\{\mathcal{F}\{u_{FLCoS}(x,y)\}\cdot H_{act}(\xi,\eta)\}, \tag{5}$$

where $u_{act}(x,y)$ is the complex hologram after active filtering, and $H_{act}$ is the content-dependent Fourier filter that blocks all areas except the region corresponding to the holographic content, as depicted in Figure 3. This active Fourier filtering serves as an additional constraint on the optimizer, forcing it to ignore all areas except the target region. In VR, where the target is typically a full-FoV image, this constraint has no significant effect on loss reduction. However, in AR scenarios where sparse virtual content is presented on a full-FoV real world, this constraint enhances the algorithm's performance significantly. Specifically, the proposed active-filter based algorithm solely focuses on clear holographic image presentation, delegating the task of managing empty areas to the physical Fourier filter. In contrast, the conventional algorithms generate both the black background and the sparse target image, imposing a significant load on SLM and resulting in lower performance. For a more detailed analysis of the algorithm's performance under sparse virtual content conditions, refer to Section 2.4. Details on creating an appropriate active Fourier filter are provided in Section S1 (Supporting Information).

After active Fourier filtering, we simulate free space propagation with Fresnel approximation as follows,

$$u_{output}^{l}(u,v)=\frac{e^{jkz_l}}{j\lambda z_l}e^{j\frac{k}{2z_l}(u^2+v^2)}\cdot\mathcal{F}\{u_{act}(x,y)\cdot e^{j\frac{k}{2z_l}(x^2+y^2)}\}. \tag{6}$$

Here, the wavenumber $k$ is defined as $k = \frac{2\pi}{\lambda}$ where λ represents the wavelength. $l$ is the index of the target depth layer, $z_l$ is the propagation distance from FLCoS to $l^{th}$ depth layer, and $u^l_{output}$ is the reconstructed complex amplitude at $l^{th}$ depth. During the Fourier transform, we set the spatial frequency as $f_x = \frac{u}{\lambda z_l}$ and $f_y = \frac{v}{\lambda z_l}$ to numerically simulate the Fresnel propagation. Note that the spatial frequency depends on the wavelength λ resulting in size differences among RGB holographic images in optical experiments. Thus, we compensate for the size mismatch between R, G, and B images by utilizing a pre-compensated target image. Finally, our loss function is,

$$\underset{u_{src}}{\operatorname{argmin}} \sum_l \left| s \frac{1}{T} \sum_t \left| u^l_{\text{output}}(u, v) \right|^2 - I^l_{\text{target}} \right|, \tag{6}$$

where $s$ is the learnable scaling factor, $I^l_{target}$ are target R, G, B images at distance $z_l$ and $T$ is the number of time-multiplexed bCGH for speckle-free holographic reconstruction. We optimized initial complex hologram $u_{src}$ using this loss function and obtained optimized binary hologram $u_{FLCoS}$.

### 2.3. Time-multiplexed system operation

Presenting a realistic AR environment through the proposed system requires two more prerequisites: per-pixel matching between the holographic content and the real-world mask, and synchronization of the FLCoS, DMD, and RGB laser. Per-pixel matching is essential for establishing immersive AR imagery, while synchronization between multiple devices is critical for time-multiplexed operation. Detailed implementations of these prerequisites are provided in Section S1 and S2 (Supporting Information). Here, we focus on the time-multiplexed operation scheme of the system, assuming that synchronization between the FLCoS, DMD, and RGB laser has been achieved and that the reconstructed holographic image is properly matched with the generated real-scene mask.

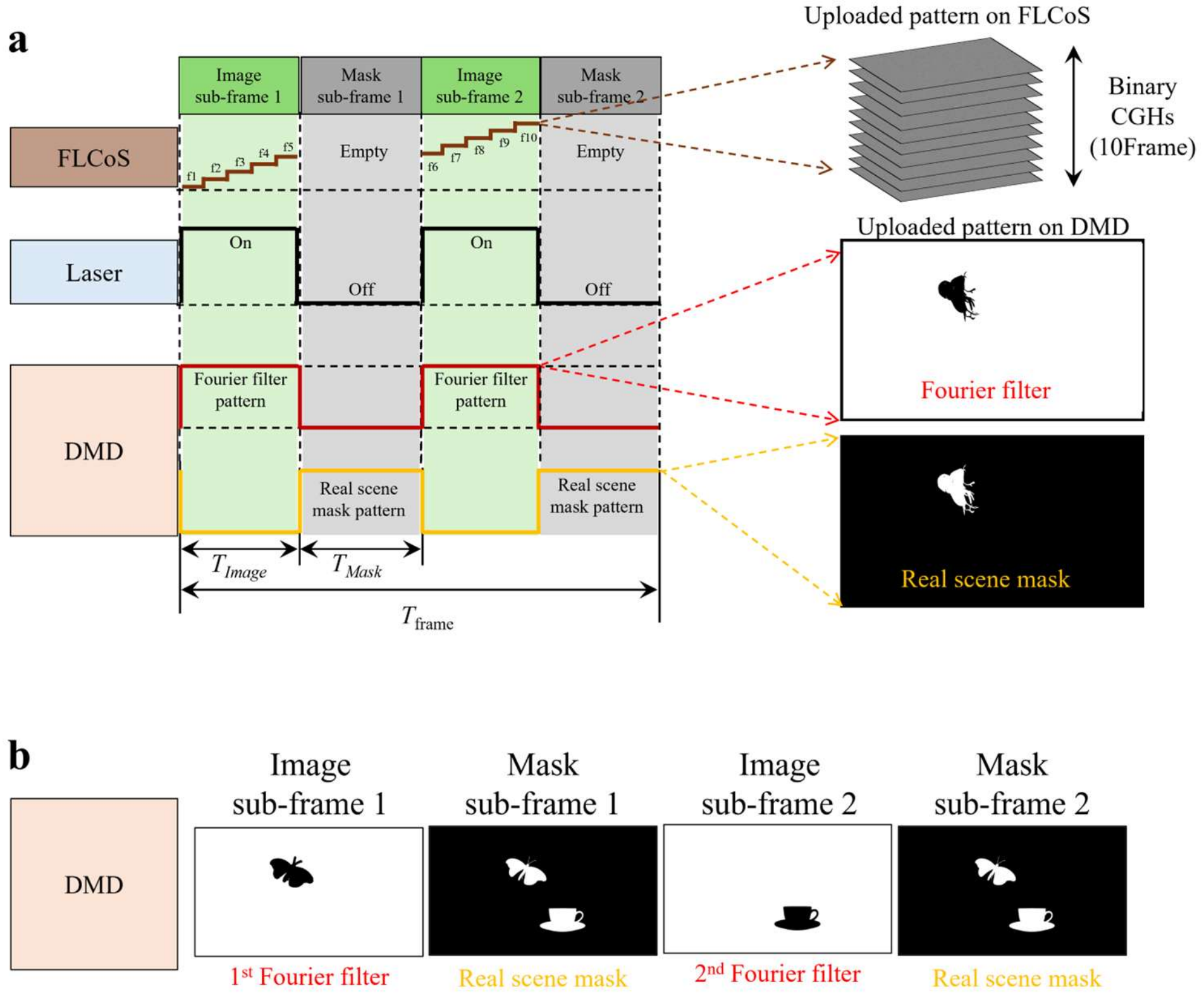

**Figure 4.** Synchronized operation of FLCoS, DMD, and laser. a) When the target image is confined to half of the image area, the DMD displays the same Fourier filter pattern in ISF 1 and 2. b) When the image spans the full area, the DMD presents two distinct patterns in ISF 1 and 2, corresponding to the upper and lower part of the target image, respectively.

**Figure 4a** illustrates the synchronized operation of the proposed system. To enable a single 4f system to function as both a Fourier filter and an occlusion mask, each device operates separately within two sub-frames: the ISF, and the MSF. In the ISF, holographic content is presented, while the selectively obscured real background is delivered to the user in the MSF. During the ISF, the high-speed FLCoS sequentially displays five bCGHs to suppress speckle noise, the fiber laser emits coherent light, and the DMD functions as a Fourier filter, passing only the desired holographic virtual content. During the MSF, the FLCoS and laser are turned off, while the DMD displays the real-scene mask pattern for occlusion. For a single frame, these two sub-frames are repeated twice, achieving speckle noise suppression through ten times multiplexed bCGHs. Note that the Fourier filter and real scene mask shown in Figure 4 correspond to the virtual holographic image positioned at optical infinity. As a result, they are inverted pairs of the binarized target image. In general cases where virtual holographic images are placed at arbitrary depths, defocus blur is applied to the Fourier filter pattern while the real scene mask remains the same, as detailed in Section S1 (Supporting Information).

Although the above time-multiplexed operation allows real scene masking and Fourier filtering through the single 4f system, holographic image presentation within a single frame is confined to a half-region (the ROI area) due to the SSB encoding in bCGH synthesis. To overcome this limitation, we synthesize the bCGHs such that the bCGHs in the first and second ISFs reconstruct holographic 3D images in different ROI areas. In specific, a holographic image reconstructed on the upper side of the DMD is presented in the first ISF, while another holographic image reconstructed on the bottom side is presented in the second ISF. To support this operation, the Fourier filters uploaded onto the DMD in the first and second ISFs are divided into two patterns corresponding to the target images on the upper and bottom sides as shown in Figure 4b. By operating the system in this manner, the proposed method fully leverages the SBP of the FLCoS for holographic reconstruction.

## 2.4. Performance analysis of the active filter-based bCGH algorithm

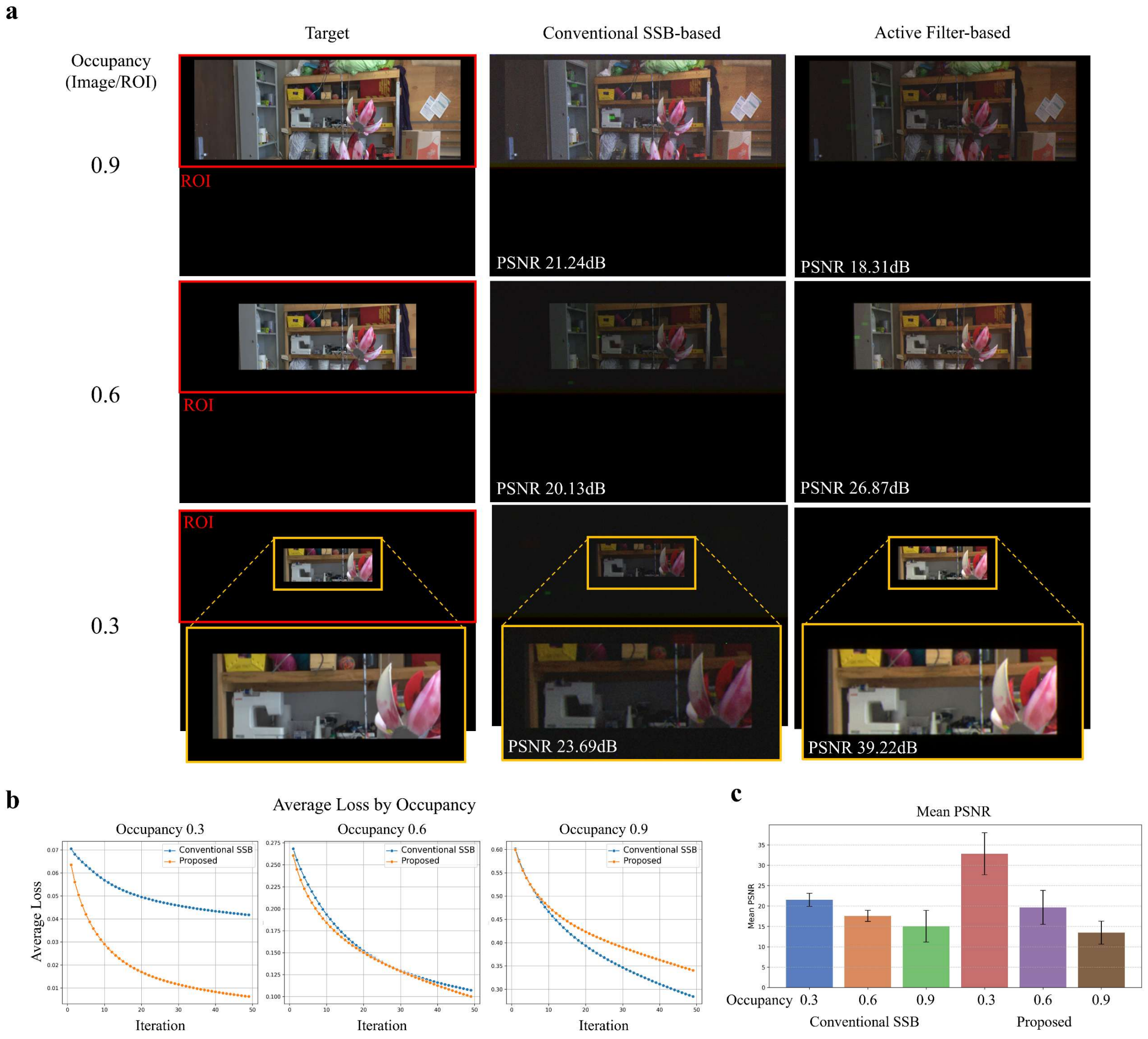

**Figure 5.** Analysis of the proposed active filter-based bCGH algorithm. a) Comparison between the conventional SSB-based bCGH algorithm and the proposed algorithm under three different occupancy conditions. b) The average loss graph. c) The mean PSNR across 24 input images. The Middlebury 2021 dataset, containing 24 stereo image sets, was used to analyze the performance of the proposed algorithm. Source image: "artroom1", Reproduced under terms of the CC-BY license.[36] Copyright 2021, by Daniel Scharstein, published with the author's permission.

To evaluate the image quality enhancement of the proposed active filter-based bCGH algorithm for sparse virtual content, we deliberately manipulated the occupancy of the target image. **Figure 5** shows the results of the performance analysis and a comparison with the conventional SSB-based bCGH algorithm. For the analysis of the algorithm, we used 24 target images from the Middlebury 2021 mobile dataset.[36] In the simulation, the resolution of the output image was set to half of the FLCoS and the number of iterations in the optimization was limited to 50 for simplicity. All other settings, such as the number of time-multiplexed bCGHs and image size pre-compensation by color, were consistent with those used in the optical experiments.

Figure 5a depicts the cropped target image ("artroom1") with the designated occupancy and the resulting holographic images generated by the conventional and proposed algorithms. Here, the occupancy refers to the ratio between the ROI area (highlighted as a red square in Figure 5a) and the image area, indicating that lower occupancy corresponds to sparser virtual content. As depicted in Figure 5a, the conventional SSB-based bCGH algorithm exhibits small variations in the peak-to-noise ratio (PSNR) when the occupancy of the target image decreases from 0.9 to 0.3. However, as the occupancy of the target image decreases, the proposed algorithm demonstrates significantly better loss reduction compared to the conventional method. Our analysis identifies an occupancy of 0.6 as the midpoint, as depicted in Figure 5b, where the performance of both methods becomes nearly identical. These findings are consistent with the mean PSNR values of the 24 output images, as shown in Figure 5c. Under low-occupancy conditions, the proposed algorithm achieves a mean PSNR of 33 dB, while the conventional SSB-based bCGH algorithm attains approximately 22 dB. This analysis demonstrates that the proposed active filter-based bCGH algorithm delivers superior image quality for sparse virtual content, which characterizes the majority of AR scenarios.

### 2.5. Experimental demonstration

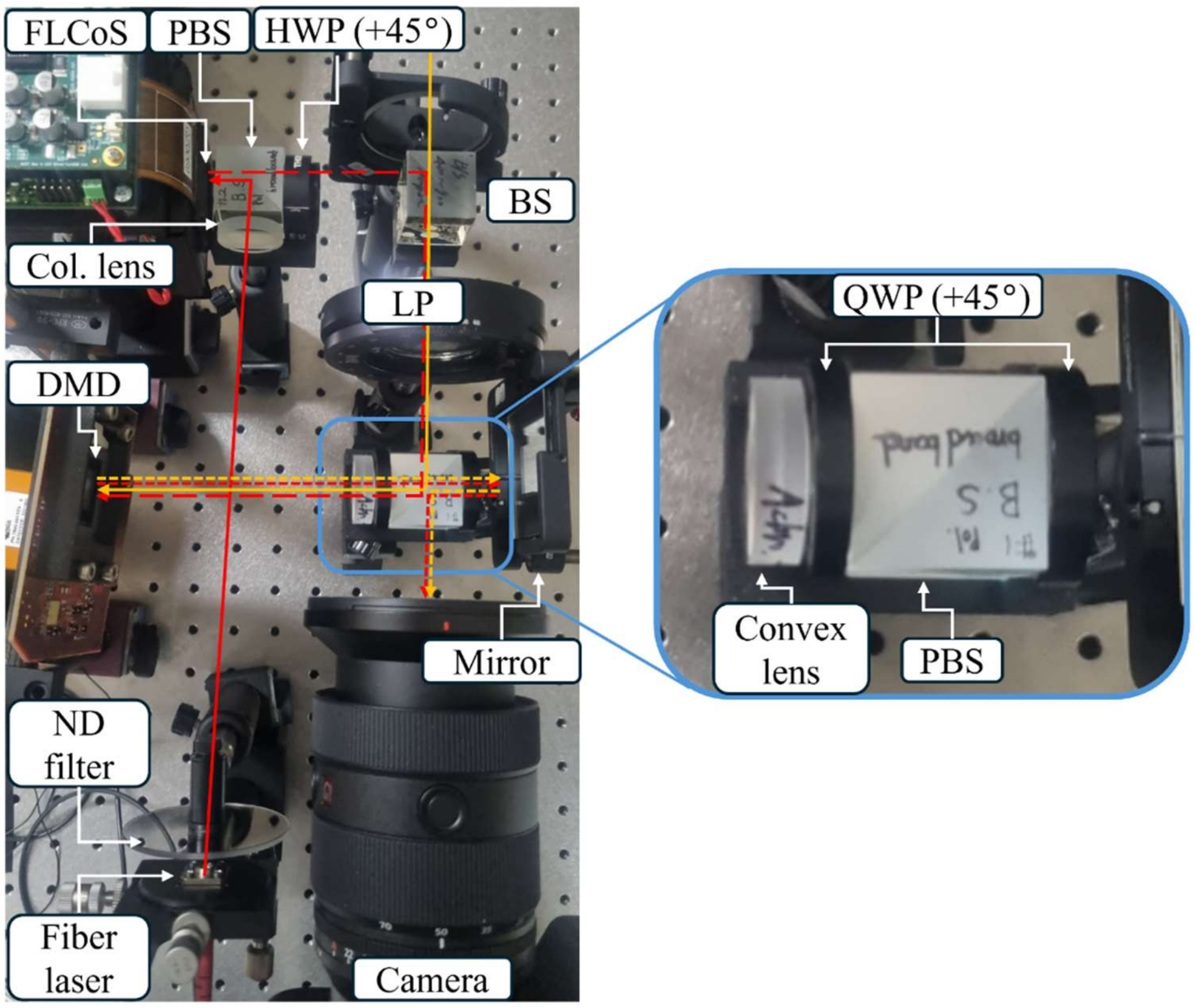

**Figure 6.** Experimental setup of the proposed system.

The proposed system was implemented as a benchtop prototype. **Figure 6** illustrates the experimental setup of the proposed system. A full-color fiber laser (FISBA, READYBeam™ Ind1) was used as the light source for the holographic display and a neutral density (ND) filter was placed in front of the laser to control the brightness of the coherent light. To provide collimated light to the holographic display, a collimation lens with a focal length of 200 mm was employed. We utilized the FLCoS (Kopin, QXGA-R10) with a resolution of 2048 x 1536 as a binary holographic display. The refresh rate of the FLCoS is 4500Hz, enabling the display of multiple bCGHs within an eye integration time. A PBS and an HWP were placed in front of the FLCoS and mounted on a custom holder with a collimation lens. A BS was positioned after the HWP to reflect the light of the holographic image towards a linear polarizer (LP). The LP filtered out the s-polarized light, allowing only the p-polarized component to pass through and enter the central optical system, highlighted by the blue rectangle in Figure 6. Here, a PBS, a convex lens and two achromatic QWPs (Edmund Optics, 46-558) were placed, constituting two folding optics. These folding optics incorporated a DMD (Texas Instruments, DLP6500) on the left and a mirror on the right. With a refresh rate of 9800 Hz, the DMD easily supports the time-multiplexed operation described in Section 2.3, and shadow casting with 8-bit depth, as explained in Section S3 (Supporting Information). Finally, a camera (SONY, A7M4) positioned below the PBS captured both the output holographic image and occluded real scene simultaneously. Note that a convex lens was additionally placed in front of the BS in our implementation to control the real object distance beyond the limits of the physical size of the optical table.

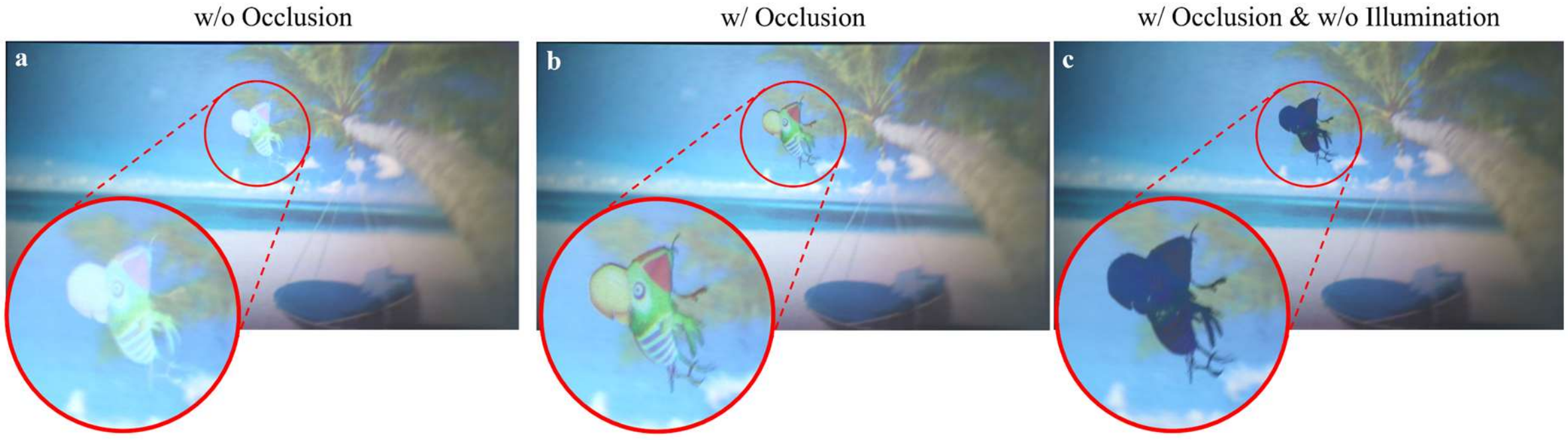

**Figure 7.** Experimental results comparing the occlusion effects on AR holographic displays. a) Conventional AR scene without occlusion. b) Occlusion-supported AR scene generated by the proposed method. c) Occlusion-supported AR scene without laser illumination. See Video 1 (Supporting Information) for a real-time video of the monochromatic version with mask-only scene. Camera settings: F/4, 1s, ISO-100.

**Figure 7** shows the experimental results comparing the difference between the conventional (occlusion-unsupported) and proposed (occlusion-supported) holographic AR scenes (See Section 4 for experimental details). In this experiment, a printed real background ("beach") was optically placed at infinity using the additional convex lens on the world side. The reconstructed image ("Pirate bird") from the time-multiplexed 10 bCGHs was positioned at a distance of 20 m. Figure 7a depicts the holographic AR scene without occlusion. Although holographic reconstruction produced a speckle-suppressed, high-quality virtual image, the real-world light significantly degraded the contrast and visibility of the virtual image, resulting in an unrealistic AR scene. In contrast, the occlusion-supported AR scene in Figure 7b demonstrates that a high-contrast holographic image can be achieved in an AR environment by masking the real-world light entering the virtual image area. Figure 7c shows the real-world mask applied to the virtual image, verifying the occlusion capability of the proposed system. Through this experiment, we confirmed that the bCGHs generated by the active filter-based algorithm produce high-quality reconstructions, and that occlusion significantly enhances the contrast and visibility of the holographic virtual image in AR scene.

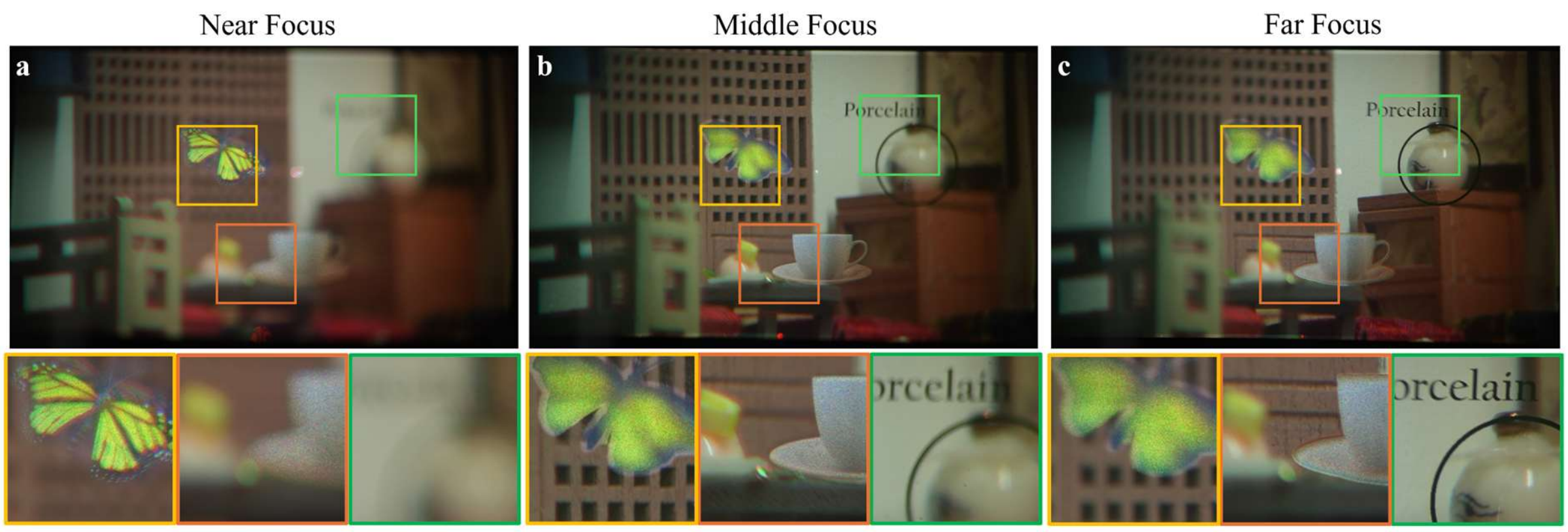

**Figure 8.** Experimental results of the proposed holographic augmented reality display with occlusion support, captured with the camera focused at a) near, b) middle, and c) far distances. A 3D real target ("Traditional house miniature") is arranged across near, middle, and far depths, while virtual holographic images ("Butterfly" and "Mug Cup") with real scene occlusion masks are positioned at the near and middle depths. At the far depth, black letters ("Porcelain") and a circle are displayed exclusively through an occlusion mask, demonstrating the system's occlusion capability.

**Figure 8** shows the 3D AR scenes generated by the proposed holographic OST-AR NED system. In this experiment, the time-multiplexing scheme introduced in Figure 4b was additionally adopted to fully leverage the SBP of the FLCoS and provide maximum FoV of the system at the expense of the occlusion ratio. (See Section 4 for experimental details). Figure 8a–c depicts the resulting 3D AR imagery produced by the proposed method. The opaque holographic virtual images (“Butterfly” and “Mug Cup”) were positioned at 3 m and 10 m depth, respectively, whereas the black letters (“Porcelain”) and a circle, created solely by the occlusion optics, were placed at optical infinity. A real 3D target (“Traditional-house miniature”) was optically located between 3 m and 20 m. When the camera was focused on the near plane as shown in Figure 8a, the opaque butterfly appeared with enhanced contrast. The mug cup, observed at the intermediate focus depicted in Figure 8b, exhibited almost identical quality to the real target. Thus, the proposed method successfully obscured real objects with virtual content, giving the impression that the virtual images were physically present. To evaluate the occlusion capability under the FoV-expanded, time-multiplexed operation, we displayed the black letters and circle solely via the occlusion mask as illustrated in Figure 8c. The results confirm that the masking capability remains sufficient to improve the contrast and visibility of the virtual images. Consequently, real-scene masking for 3D holographic images was successfully demonstrated, effectively reproducing the optical interactions between real objects and virtual imagery. Additional experimental results are provided in Section S4 (Supporting Information).

Various phenomena in nature result from the optical interaction between objects. A shadow is one such phenomenon, originating from the interaction between a light source and the objects. In this study, we attempted to cast a shadow by leveraging the occlusion, as depicted in **Figure 9**. To generate the shadow, we developed a Poisson distribution-based point spread function tailored for shadow rendering and convolved it with a hard-edge shadow to produce natural soft-edge shadow. After obtaining the 8-bit grayscale soft-edge shadow image, the DMD displayed the shadow image with real scene mask in the MSF. Since the DMD is a binary-amplitude SLM, pulse-width modulation (PWM) operation was employed. See Section S3 (Supporting Information) for detailed methods of shadow generation and operation.

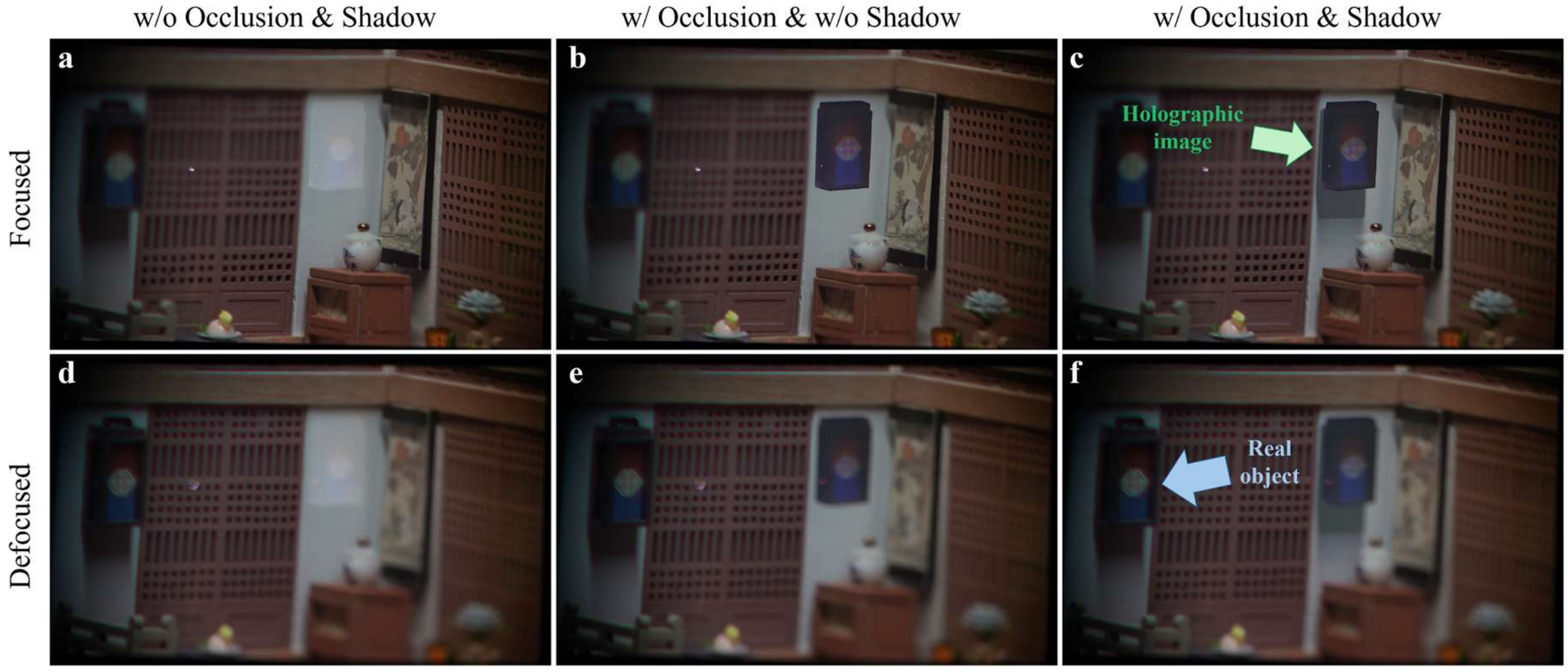

**Figure 9.** Experimental results of casting the shadow of the virtual image ("Traditional lamp"). For comparison, the AR scenes were captured with a–c) the virtual image in focus and d–f) the virtual image out of focus. a,d) Conventional AR holographic scenes are shown in the first column, b,e) occlusion-supported AR scenes in the second column, and c,f) AR scenes with both occlusion and shadow in the third column.

In Figure 9, the holographic virtual image ("Traditional lamp") was placed at a depth of 20 m, and the real object ("Traditional-house miniature") was optically positioned in the depth range from 5 m to 20 m. The experimental results in the first row and the second row were captured with the camera focused at 20 m and 5 m depth, respectively. As illustrated in Figure 9a,d, the conventional holographic AR scenes present the virtual image with low visibility and contrast. Here, the degradation caused by real-world light was more severe compared to the previous results in Figure 7, as the virtual content was overlaid on a white background. After employing occlusion, visibility and contrast were significantly improved as shown in Figure 9b,e. However, the AR imagery remained unrealistic due to the absence of a shadow for the virtual image. Figure 9c,f show the final AR imagery with a real scene mask and shadow. These results clearly demonstrate that immersive and realistic AR scenes can be achieved by supporting occlusion and casting shadows for the virtual image.

**3. Discussion and Conclusion**

In summary, we propose an occlusion-capable holographic AR display and an active Fourier filter-based algorithm for bCGH synthesis. Leveraging the structural similarity between occlusion and Fourier filtering optics, our system integrates these functionalities using a single DMD within the folding structure, thereby reducing the overall system size. The proposed algorithm based on the active Fourier filtering achieves superior performance in reconstructing localized virtual content. In optical experiments, we demonstrated that the speckle-suppressed, high quality virtual images can be generated using bCGHs synthesized through our algorithm. By supporting the occlusion, the depth order, visibility, and contrast of the holographic image in an AR scene were noticeably enhanced, enabling the realistic AR presentations. Realism in the AR environment was further improved by incorporating shadow casting. However, our system has potential for further improvement in several aspects.

We introduced occlusion as a tool for imitating the optical interaction between the real scene and the virtual holographic image. By handling occlusion and generating shadows, the realism in AR imagery was successfully enhanced. Nevertheless, there remains potential for further enhancement. For example, the interaction with a light source is not limited to shadow casting. It also affects the brightness of objects, creating relatively bright and dim areas depending on their distance to the light source. Similarly, the interaction between real objects and virtual objects results in not only occlusion but also other effects based on the material properties of the foreground objects. For instance, if a virtual half-mirror is introduced in the foreground, the light from the real scene in the background should remain observable, while simultaneously generating reflected light from the other side of the real world. Creating a realistic AR scene requires consideration of additional aspects related to the optical interaction as well, such as texture, light scattering, and so on. We leave these prospective improvements for future work.

In the current prototype, developed using off-the-shelf components, our primary objective was to demonstrate the feasibility of the proposed system. For experimental convenience, the eyebox was expanded to approximately 10mm at an eye relief of 20mm;

however, this configuration resulted in a see-through FoV of around 9° and a virtual image FoV of about 4°. These limitations are largely attributed to the lens focal lengths of the lenses and the restricted SBP of the employed FLCoS device. Nevertheless, the proposed architecture is inherently scalable. By incorporating high-NA metalenses[37] and leveraging high-order terms, we expect the achievable see-through FOV to exceed 50°, and the virtual image FOV to surpass 10°, while simultaneously reducing system size. Furthermore, the integration of a dynamic deflector [38-40] has the potential to further expand the see-through and virtual image FoVs to approximately 70° and 30°, respectively, albeit at the cost of additional time-multiplexing.

In our implementation, we employed the DMD for both occlusion and active Fourier filtering, as the DMD board offers advantages in precise synchronization. However, the DMD we adopted is a diagonally slanted mirror array, providing two states with angles of +12°, and −12°. To compensate for this inherent slant, the DMD panel was globally tilted along both the vertical and horizontal axes. While this global tilt allowed the incident light to reflect back towards the convex lens, the slanted panel introduced several issues in our system. One issue is the depth variation depending on the spatial position. Since the DMD is slanted, the distance to the convex lens varies with the spatial position, causing depth shifts in both the real scene and the virtual image. Another issue is a chromatic aberration of the real scene created by the diffraction from pixelated structure of the DMD panel, as observed around the UFO object in Figure S10 (Supporting Information). Although the diffraction fundamentally arises from the pixelated structure of the DMD, the global tilt can amplify the separation between desired light and diffracted light. To address this, the DMD can be replaced with a standard flat-panel SLM such as amplitude-only LCoS, which has not been used in our implementation due to the difficulty in precise synchronization with other devices. We believe that the synchronization with a standard flat-panel SLM can be achieved through advanced circuit design.

The proposed system adopted a DMD panel fixed at the Fourier plane, providing a 2D occlusion mask at optical infinity. As a result, the system delivered precise hard-edge masking for virtual images located at infinity, while relatively blurred masks were applied to virtual images positioned at closer depths. Although we calculated the appropriate mask pattern for such depth-mismatched cases as explained in Section S1 (Supporting Information), ideal occlusion is only achievable through depth matching between the virtual content and the corresponding mask, which requires 3D occlusion. A recent study on varifocal occlusion using an LC-based varifocal lens demonstrated that folding optics-based 3D occlusion is feasible by optically manipulating the depth of the mask while preserving the real scene depth[6]. This approach could be directly applied to our work and contribute to a more realistic AR presentation. Furthermore, various schemes for 3D optics, including light field, varifocal, and multifocal displays, hold potential for enabling 3D occlusion. We anticipate that advancements in 3D occlusion will broaden the applicability of the proposed method.

The proposed design utilized folded 4f optics for occlusion and Fourier filtering. However, after passing through the 4f system, the real scene was spatially inverted. Consequently, the experimental results presented in this paper were rotated by 180° after captured by the camera to compensate for the inverted real scene. This spatial inversion can be optically corrected by employing an additional 4f optical system or a Schmidt–Pechan prism, as reported in previous studies[15]. Another issue lies in the depth shifting of the real scene, commonly referred to as pupil matching problem. In our setup, the light from the real scene propagated through an additional optical path after being relayed by the 4f optics. Consequently, the depth of the real scene was shifted farther by an amount corresponding to the additional propagation distance. Due to the compact nature of the system, the additional propagation distance was only approximately 5 cm in our implementation, which can be ignored in most cases. More fundamental remedy to this pupil mismatching issue would also be possible by applying the methods reported in previous studies.[41]

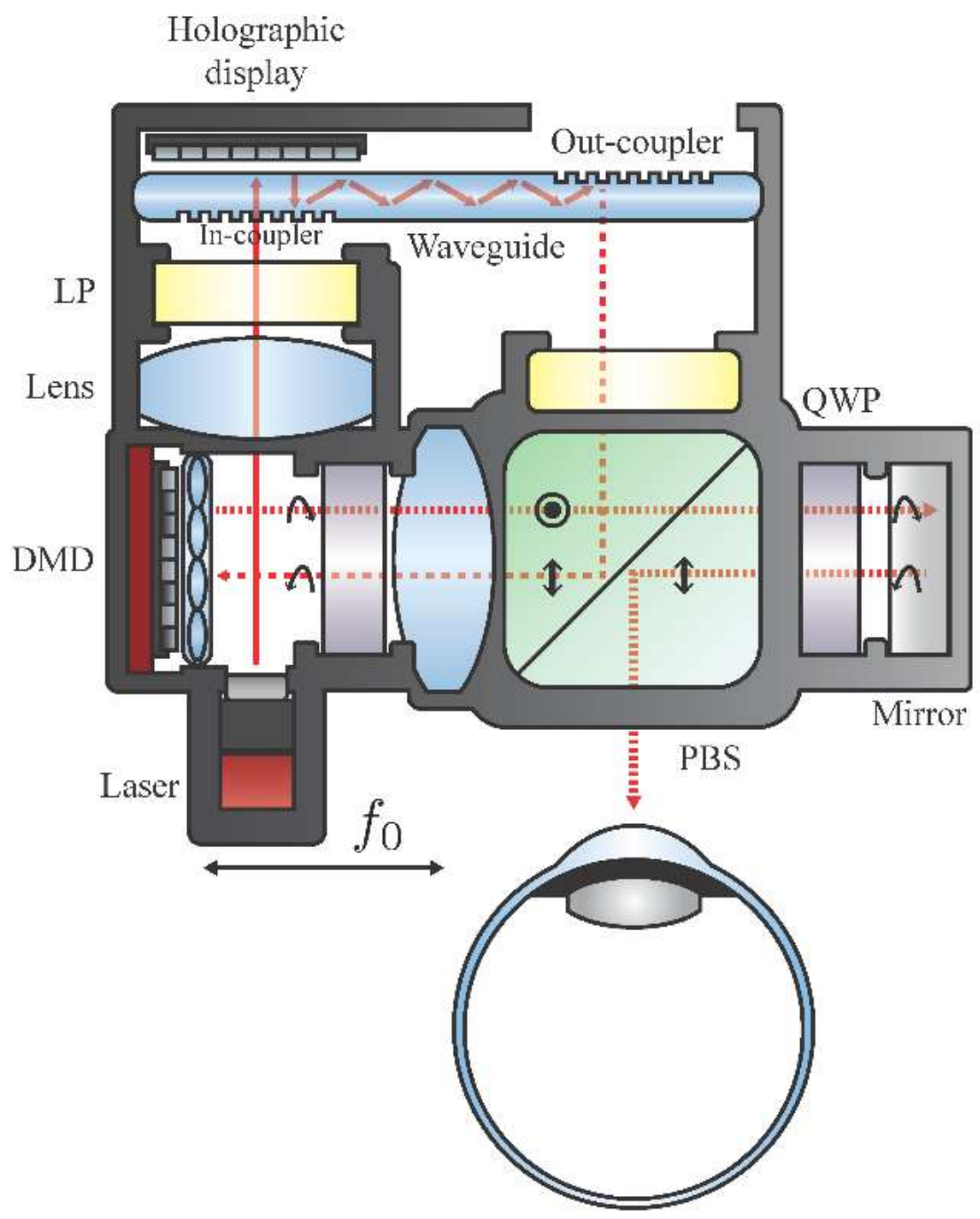

**Figure 10.** Compact version of the proposed system.

In the proposed system, Fourier filtering and real scene masking were achieved using a single DMD within a folded 4f system. This design reduced the number of optical components and devices by half. Nevertheless, there is still potential to compress the overall system size further.

**Figure 10** illustrates a schematic of a compact version of the proposed method, which includes an additional micro lens array (MLA) placed in front of the DMD. To reduce the overall size, the bulky combiner optics, including PBS1, HWP, and BS, can be replaced with slim waveguide combiner optics. Considering recent advancements in waveguide holographic AR displays,[42, 43] our system can benefit significantly from this approach. For more compact form factors, a convex lens with a smaller focal length can be employed, reducing the gap between components. The MLA placed close to the DMD enables light field occlusion,[44, 45] allowing the real scene mask to be placed at an arbitrary depth without sacrificing form factor. Additionally, the HWP is omitted in the compact version, assuming that the holographic display uses a phase-only LCoS. Since the proposed occlusion-enabling architecture supports both amplitude- and phase-only holographic displays, it offers flexibility in selecting compact devices.

## 4. Experimental Section

In the experiment of Figure 7 and 9, the time duration of the ISF, $T_{\mathrm{image}}$, and MSF, $T_{\mathrm{mask}}$, was set to 2.4ms and 14.3ms respectively. As a result, theoretical occlusion ratio of the mask was approximately 85.6% indicating that the masking area blocked 85.6% of the real-world light. The measured occlusion ratio in the experiments was 76.4% due to the stray light from the real world and switching time between the frames. The total duration of a single frame, $T_{frame}$, was 33.3ms which corresponds to 30Hz operation. For full-color presentation, we captured R, G, B monochromatic images and combined them digitally. In occlusion supported case, the DMD displayed the Fourier filter pattern in ISF and the real scene mask pattern in MSF, as depicted in Figure 4a. To present occlusion unsupported AR scene, the DMD displayed the black frame on every MSF. This modified operation only reduced the occlusion ratio from 76.4% to 0% while keeping all other operation parameters, including the intensity of the light source, unchanged. This specific DMD operation of generating AR scene without occlusion, was used again in following experiments for comparison with proposed method.

In the experiment of Figure 8, we applied the FoV-expanded operation scheme in Figure 4b. To maintain 30Hz operation and keep the number of bCGHs at 10 for each image, we adjusted $T_{Mask}$ and repeated the DMD operation in Figure 4b twice. $T_{Mask}$ was set to 5.9ms for both MSF 1 and MSF 2, achieving 60Hz operation for one cycle ($2T_{image} + 2T_{Mask} = 2\times2.4\text{ms} + 2\times5.9\text{ms} = 16.6\text{ms}$). Since holographic images in each ROI were reconstructed using 5 bCGHs per cycle, we set a single frame consisted of two cycles. While the DMD patterns were repeated across the two cycles, the 10 bCGHs for each ROI were divided into two sets and displayed during their respective cycles. With this operation, the occlusion ratio in Fig. 5 was theoretically around 71% and measured to be 58.2%. Note that the occlusion ratio can be increased by reducing the time duration of the ISF, $T_{image}$. In our implementation, however, it was kept the same as in the previous case due to the limited refresh rate of the FLCoS.

**Supporting Information**

Supporting Information is available from the Wiley Online Library or from the author.

**Acknowledgements**

This research was supported by the National Research Foundation of Korea (NRF) grants funded by the Korean government (MSIT) [Grant No.RS-2022-NR070432, 34%]; [Grant No. RS-2024-00416272, 33%]; and [Grant No. RS-2024-00414230, 33%].

**Conflict of Interest**

The authors declare no competing interests.

**Author Contributions**

J.-H.P. planned and supervised the project. J.-H.P. and W.H. conceived the original idea and designed the system along with the corresponding algorithm. W.H. and C.L. performed the simulations and carried out the detailed experiments. All authors discussed the results and contributed to writing the manuscript.

**Data Availability Statement**

The data that support the findings of this study are available from the corresponding author upon reasonable request.